\documentclass[journal,article,accept,pdftex,moreauthors]{Definitions/mdpi} 
\firstpage{1} 
\pubvolume{1}
\issuenum{1}
\articlenumber{1}
\pubyear{2025}
\copyrightyear{2025}
\datereceived{ } 
\daterevised{ } 
\dateaccepted{ } 
\datepublished{ } 
\hreflink{https://doi.org/} 
\pdfoutput=1 

\Title{Can Invisible Psychological Traits Organize Visible Network Structure?
A Complex Network Analysis of Myers–Briggs Type Indicator-Based Interaction Patterns in Anonymous Social Networks}

\TitleCitation{Title}

\Author{Seyed Moein Ayyoubzadeh $^{1,\dagger}$ \orcidA{} , Kourosh Shahnazari $^{1,\dagger}$ \orcidB , Mohammadamin Fazli  $^{1}$  , Mohammadali Keshtparvar $^{2}$}

\AuthorNames{Seyed Moein Ayyoubzadeh, Kourosh Shahnazari}

\isAPAStyle{%
       \AuthorCitation{Lastname, F., Lastname, F., \& Lastname, F.}
         }{%
        \isChicagoStyle{%
        \AuthorCitation{Lastname, Firstname, Firstname Lastname, and Firstname Lastname.}
        }{
        \AuthorCitation{Lastname, F.; Lastname, F.; Lastname, F.}
        }
}

\address{%
$^{1}$ \quad Sharif University of Technology\\$^{2}$ \quad Amirkabir University of Technology
}

\corres{Correspondence: smoein.ayyoubzadeh16@sharif.edu (S.M.A.); kourosh@null.net (K.S.)}

\firstnote{These authors contributed equally to this work.}

\abstract{
Exploration of the impact of personality traits on social interactions within anonymous online communities poses a great challenge at the interface of networked social sciences and psychology. We analyze here whether Myers-Briggs Type Indicator (MBTI) personality types have any impact on the dynamics of interactions on a well-known anonymous chat system with over 288,000 messages from 6,076 users. Based on a data set including 940 users voluntarily providing MBTI typing and gender, we create a weighted undirected network and apply a range of network-science measures—such as assortativity, centrality measures, clustering coefficients, and community detection with the Louvain algorithm—to estimate the level of personality-based homophily and heterophily. Contrary to previous observations in structured social settings, our research exhibits a dominance of heterophilous interactions (89.3\%), particularly among cognitively complementary types, i.e., NT (Intuitive-Thinking) and NF (Intuitive-Feeling). However, there is still a moderate level of personality-based homophily (10.7\%), notably among introverted intuitive personalities (e.g., INTJ, INFP, INFJ), reflecting an underlying cognitive alignment that takes place regardless of identity markers. The interaction network exhibits scale-free properties (with a power-law exponent \( \alpha = 1.45 \)), with high-degree hubs supporting communication among various personality types. Notably, individuals with intermediate MBTI categories often serve as bridging personalities, having high centrality scores and fostering community cohesion. In contrast, gender is a stronger homophily attribute, as evidenced through much stronger levels of female users' group interactions compared with male users' group interactions. While MBTI type influences minor preferences for interactions, community structure is low on modularity (\( Q = 0.2584 \)) with dynamics being influenced mostly through loose patterns of conversations and not through following types strictly. The findings show that, in the absence of identity cues, psychological traits subtly shape online behavior, combining exploratory heterophily with subtle homophilic inclinations. This study offers a generalizable framework for the integration of personality theory with cutting-edge network analysis and implies important implications for the design of social media platforms that are both personality-sensitive and privacy-aware.
}

\fancypagestyle{plain}{
  \fancyhf{} 
}

\begin{document}


\section{Introduction}

\subsection{The Psychology of Personality and Social Interaction}

Long central to psychology, personality research has evolved from different theoretical models explaining how individual variations affect cognitive, emotional, and behavioral processes. Among these systems, the \textit{Myers-Briggs Type Indicator} has become rather well-known for its methodical classification of people into 16 personality types. Myers-Briggs Type Indicator dimensions seek to capture basic differences in human cognition and interaction preferences by developing from Carl Jung's theory of psychological types. Although psychometric restrictions continue to cause controversy in some scientific circles about the Myers-Briggs Type Indicator, its extensive application in organizational psychology, behavior prediction, and social interaction studies emphasizes its relevance in personality research.

The development of social networks is one of the main ways personality influences human interaction. Research spanning decades has shown that people do not create relationships haphazardly; rather, they show observable tendencies in which they associate, somewhat motivated by personality compatibility (or incompatibility).\citep{mcpherson2001birds, charilaos2014analysis, gilal2017software}. Particularly interesting is the function of homophily in social interactions: people are more likely to create bonds with others who have similar characteristics, causing network clustering and more frequent communication inside homogeneous subgroups.

\subsection{Homophily and Heterophily in Social Networks}
Emphasizing that ``similarity breeds connection,'' the basic concept in sociology and network science is \textit{homophily}—coined by Lazarsfeld and Merton in 1954 \citep{lazarsfeld1954friendship}. Seen in many spheres—race, religion, education, and personality traits among others—this phenomenon has been widely documented. In personality-based homophily, people with parallel cognitive styles, emotional processes, or interaction preferences build closer social ties, thus supporting shared beliefs and behavior standards \citep{park2015structural}.

Conversely, studies on \textit{heterophily} suggest that interaction between different people—typically complementary in nature—can enhance information variety and network efficiency \citep{granovetter1973strength, mcpherson2001birds}. Cross-type Myers-Briggs Type Indicator interactions (such as the classic Extraversion–Introversion or Thinking–Feeling pairings) challenge homogeneous thinking within closed clusters and bring variation in cognitive perspectives.

While homophily dominates in structured environments such as workplaces, classrooms, and social media platforms, the degree to which it persists in anonymous networks remains an open question. 

\subsection{Theoretical Implications of Anonymous Interaction Networks}

Online anonymity removes crucial markers of identity including demographic information, social capital, and physical appearance \citep{christopherson2007conceptual}. This raises a basic research issue: \textit{Does homophily persist when identity cues are eliminated, or does anonymity generate heterophily by lowering social constraint?} Different theories produce different predictions:

\begin{enumerate}
    \item \textbf{Homophily persistence hypothesis:} Depending on personality signals buried in response timing, preferred subjects, and communication style, like-minded users can even be found under anonymity.
    \item \textbf{Heterophily amplification hypothesis:} Consistent with the ``reduced social cue'' theory \citep{sproull1986reduced}, anonymous settings could motivate people to interact with different personalities, especially if conventional barriers (e.g., shyness, social anxiety) are removed.
\end{enumerate}

Although theoretically significant, empirical studies on anonymous chat behavior in connection to Myers-Briggs Type Indicator are still rare.

\subsection{Empirical Studies on Myers-Briggs Type Indicator in Network Science}
Myers-Briggs Type Indicator research on network interactions has investigated personality clustering in controlled groups—that is, teams, classrooms—but results have been mixed:
\begin{itemize}
    \item \textbf{Homophily findings:} Some research shows unequivocally that similar Myers-Briggs Type Indicator types cluster together in work teams and friendship networks, thus supporting self-reinforcing cognitive patterns \citep{gilal2017software, yan2022friendship}.
    \item \textbf{Heterophily benefits:} Some personality traits enhance group outcomes—such as the inclusion of Extraverts and Introverts optimizing team discussion, or Thinking-Judging types ensuring goal completion \citep{garousi2018team}.
\end{itemize}
These studies, meanwhile, have focused on non-anonymous, identity-based platforms. Myers-Briggs Type Indicator-based network building in essentially anonymous digital environments is not the subject of much large-scale study.

\subsection{Core Research Questions and Hypotheses}
By addressing the following research questions, this work seeks to close current gaps:

\begin{enumerate}
    \item \textbf{Does homophily driven by Myers-Briggs Type Indicator persist in anonymous chat systems?} We expect to see high assortativity in interaction graphs if personality-based homophily spreads into environments free of identity.
    
    \item \textbf{Do particular Myers-Briggs Type Indicator pairings correlate with higher interaction intensity?} If heterophily promotes cross-type communication, different Myers-Briggs Type Indicator combinations—like Extraversion–Introversion or Thinking–Feeling—may show increasing engagement over time.
    
    \item \textbf{In Myers-Briggs Type Indicator-driven online networks, what structural patterns show up?} Analyzing connectivity structures and clustering coefficients helps us to assess whether digital Myers-Briggs Type Indicator-driven interactions reflect actual personality clustering patterns.
\end{enumerate}

\subsection{Novel Contributions}

This research addresses several critical gaps:
\begin{itemize}
    \item Investigating \textbf{pure personality effects} on network formation without external social 
    \item factors
    \item Analyzing \textbf{Myers-Briggs Type Indicator-driven anonymous digital interactions} at scale.
    \item Providing \textbf{network science metrics} to quantify homophily, heterophily, and structural stability.
\end{itemize}
The outcome will enhance theoretical models of personality-driven network evolution and inform future designs for digital social platforms.

\section{Related Work}

Social network analysis, computational psychology, and online behavioral research have increasingly focused on the study of personality-driven network creation—especially with relation to the Myers-Briggs Type Indicator. Myers-Briggs Type Indicator affects homophily—the desire for similar connections—as well as heterophily—the desire for different connections—and network structure metrics like modularity, centrality, and clustering coefficient, researchers have already investigated. While important conclusions have been established in controlled environments such as offices and classrooms, knowledge of these dynamics in anonymous digital platforms—where identity cues beyond personality are mostly lacking—remains much lacking. This section aggregates earlier studies on Myers-Briggs Type Indicator effects on homophily and heterophily patterns, personality in digital and anonymous environments, and research approaches applied in personality-based network studies.

\subsection{Myers-Briggs Type Indicator-Driven Homophily and Heterophily in Social Network Formation}

\subsubsection{Evidence for Homophily in Myers-Briggs Type Indicator-Based Networks}
Long a basic concept in social network theory, \textit{homophily}—where people preferably interact with others with similar traits—has been influenced by Myers-Briggs Type Indicator, showing homophilic clustering patterns in many studies on personality-based socializing.

\citeauthor{gilal2017software} studied Myers-Briggs Type Indicator-based team building in software development environments by means of network metrics including betweenness centrality and degree centrality in order to identify key personality-driven nodes. They discovered that ISTJ personalities became central players, thus supporting the hypothesis that various personality types show preferences for related cognitive styles in organized cooperation. Likewise, \citeauthor{charilaos2014networking} sought to match Myers-Briggs Type Indicator-driven homophily inside social media friendships with Facebook network clustering patterns. Though the small size of the dataset they used limits their results, they did find a link between personality similarity and shorter network paths, which implies personality-based affinity.

Apart from social and professional venues, \citeauthor{varona2020assessing} devised a mathematical model to evaluate Myers-Briggs Type Indicator-based role alignment. The researchers discovered that team task assignment mostly depends on personality-based preferences. This outcome supports the homophily theory, according to which personality traits define group dynamics.

\subsubsection{Heterophilic Complementarity in Network Connectivity}
Strong arguments for homophily notwithstanding, many studies have shown that heterophilic personality connections—where people with different traits are linked—make networks more efficient and connected.

\citeauthor{garousi2018team} examined, empirically, how team compositions set around the Introversion–Extraversion Myers-Briggs Type Indicator dichotomy affected problem-solving effectiveness. Their data showed that mixed-type teams did better than homogeneous teams on projects. Surprisingly, though, team cohesiveness remained rather constant. This suggests an interaction and performance trade-off whereby cognitive diversity improves results but not always social bonding.

\citeauthor{yan2022friendship} further applied exponential random graph modeling (ERGM) to assess personality complementarity in friendship networks. They found, against the homophily assumption, that complementarity rather than similarity drove personality-driven clustering. According to the theory of network efficiency, heterophilic ties act as bridges enabling different kinds of information to pass across the network more readily and so reducing the number of broken links.

\subsection{Personality and Anonymous Online Interaction}

\subsubsection{Challenges in Mapping Myers-Briggs Type Indicator Homophily in Anonymous Platforms}
Sometimes conventional research on personality-based networking depends on well defined social structures where stability and repetition support homophilic tendencies, such as social media interactions or corporate teams. High-churn, completely anonymous interactions present a unique challenge since personality expression is less clear and leads to random network behavior.

\citeauthor{kang2017homophily} looked at anonymous online discussion forums to find whether personality homophily persisted when users lacked obviously identifiable demographic traits including age or location. Their studies disproved earlier hypotheses by demonstrating that whereas personality similarity had little effect on exchange frequency, socioeconomic and educational similarity showed strong homophilic tendencies. The results suggest that anonymity reduces traditional personality-based clustering systems—knowledge vital for anonymous chat platforms where users interact without continuous identities.

\citeauthor{jiao2024social} conducted a more overall qualitative study on how Generation Z combines Myers-Briggs Type Indicator as a social heuristic in their interactions across several platforms. Myers-Briggs Type Indicator was found to be both a social ``bridge'' and a ``barrier,'' depending on context. Thus, personality-based interaction may not be essentially stable across platforms but is rather shaped by social conventions and platform affordances.

\subsubsection{Personality Manifestation in Computer-Mediated Communication}
Personality influences online behavior differently depending on communication modalities. Research has shown that synchronous interactions (e.g., real-time messaging) and asynchronous interactions (e.g., forum replies, delayed chat) elicit different network characteristics based on cognitive processing demands.

\citeauthor{thatcher2003smallgroup} They compared face-to-face discussions with computer-mediated communication (CMC) decision-making, examining how Myers-Briggs Type Indicator traits influenced interaction patterns in each modality. Their findings suggested that personality’s predictive power was stronger in face-to-face conversation than in text-based CMC, possibly due to the absence of nonverbal cues. This casts doubt on whether Myers-Briggs Type Indicator effects translate consistently across digital environments, particularly in anonymous settings where personality-based cues may be less pronounced.

\subsection{Network Analysis Techniques Applied to Personality-Based Social Structures}

\subsubsection{Centrality, Clustering, and Modularity in Myers-Briggs Type Indicator Networks}
Network analysis techniques are being used more and more on personality-based data to learn about structural features like modularity, betweenness centrality, and clustering coefficients.

\citeauthor{yan2022friendship} used ERGM models to look into how personality complementarity affects network clustering. It was emphasized that preference-driven connectivity can't always be taken for granted; preferences must be backed up by numbers using probabilistic modeling. This aligns with the findings of \citeauthor{perugini2015network}, who introduced a theoretical framework for conceptualizing personality as an emergent network property rather than a static latent trait.

Additionally, \citeauthor{gilal2017software} provided one of the few empirical applications of complex network analysis to Myers-Briggs Type Indicator data in team environments, using weighted interactions and network centrality measures to evaluate the relational structure of professional teams. \citeauthor{charilaos2014networking} further explored social structure dynamics within online platforms using social media network measures; however, their dataset was limited, reducing the statistical reliability of findings.

\subsection{Gaps and Future Research Considerations}

Myers-Briggs Type Indicator still has significant knowledge gaps even if linking it to network development has made great progress possible.

Most studies offer static snapshots of network creation rather than looking at how Myers-Briggs Type Indicator-based structures change. Greater understanding of network persistence trends would come from dynamic tracking techniques.

\textbf{Lack of Large-Scale Anonymous Platform Data:} Studies specifically targeted on Myers-Briggs Type Indicator in high-anonymity environments remain rare even if research by \citeauthor{kang2017homophily} questions conventional homophily assumptions.

\textbf{Absence of Asynchronous/Synchronous Communication Comparisons:} Myers-Briggs Type Indicator-driven homophilic or heterophilic networking remains unexplored under delayed vs. real-time interactions.

If these gaps are closed, next studies could enable us to better understand how personality features impact anonymous digital social structures. This might result in the development of more sophisticated prediction models or improved means of interaction in totally anonymous network environments.

In essence, past studies clarify how Myers-Briggs Type Indicator influences network development; but we need fresh approaches combining behavioral tracking with longitudinal social network analysis to ensure that the results remain valid in a wide spectrum of digital and anonymous environments.


\section{Methodology}

\subsection{Data Collection and Preparation}

\subsubsection{Description of Dataset Sources}

Anonymized chat interactions acquired via a Telegram-based anonymous social network—which lets users connect randomly without disclosing their actual identities—make up the dataset used in this work. Over six months, the data was gathered capturing a total of 288,616 messages sent among 6,076 distinct users. Among these, 940 users supplied both their Myers-Briggs Type Indicator personality type and gender, so adding to 161,753 messages in which both interacting parties had recorded Myers-Briggs Type Indicator and gender information.

Every interaction consists of timestamps, user IDs (hashed to guarantee anonymity), and conversation sessions among other metadata. Under conditions of anonymity, this dataset allows a strong analysis of personality-driven social interactions and network building.

Every data collecting process followed ethical research guidelines, so guaranteeing user privacy and institutional review board (IRB) compliance. Myers-Briggs Type Indicator's voluntary character and gender disclosure let users keep control over their data, so matching privacy-preserving guidelines.

\subsubsection{Preprocessing Steps}

Before performing network analysis, extensive preprocessing steps were conducted to clean and structure the dataset for analysis. The preprocessing pipeline consisted of the following steps:

\paragraph{Data Cleaning and Normalization}
\begin{itemize}
    \item Removed duplicate messages and system-generated notifications to maintain data integrity.
    \item Converted user identifiers into anonymized hash values to prevent traceability while maintaining unique user tracking.
\end{itemize}

\paragraph{Handling Missing and Incomplete Data}
\begin{itemize}
    \item Identified and removed users with incomplete or inconsistent Myers-Briggs Type Indicator self-reports.
    \item Excluded interactions where Myers-Briggs Type Indicator or gender data was missing for either participant.
    \item Applied text processing techniques to remove special characters, URLs, and stop words to facilitate message analysis.
\end{itemize}

\paragraph{Merging and Structuring the Dataset}
\begin{itemize}
    \item Constructed interaction logs by merging conversation data with user personality attributes.
    \item Formatted data into a structured relational schema consisting of three primary tables:
          \begin{enumerate}
              \item \textbf{Users Table}: Containing hashed user IDs, Myers-Briggs Type Indicator personality types, and gender information.
              \item \textbf{Messages Table}: Storing message content, timestamps, and sender-receiver identifiers.
              \item \textbf{Interactions Table}: Representing message exchanges between users, serving as the foundation for network construction.
          \end{enumerate}
    \item Validated the consistency of merged data by performing cross-checks on user activity timestamps.
\end{itemize}

\paragraph{Anonymization and Ethical Considerations}
\begin{itemize}
    \item Removed personally identifiable information (PII) by replacing usernames with unique hash values.
    \item Ensured all dataset processing conformed to ethical research guidelines and data protection laws.
\end{itemize}

After these preprocessing stages, the dataset was organized such that additional network analysis and personality-driven interaction modeling could be facilitated. Building the interactive network to investigate Myers-Briggs Type Indicator-driven homophily and heterophily patterns in anonymous digital environments came next.

\subsection{Network Construction}

\subsubsection{Defining Nodes and Edges}
The foundation of the interaction network is user talks collected from an anonymous Telegram social network form. Nodes are the users of this system; interactions among them define the edges. Every edge in the network stands for a chat between two users; the edge weight shows the message count in that exchange.

Information provided by users shapes the features of nodes. Every node makes up:

\begin{itemize}
    \item \textbf{Myers-Briggs Type Indicator:} User self-reported Myers–Briggs Type Indicator
    \item \textbf{Gender:} The stated gender of the user
\end{itemize}

Edges log the social connection interaction frequency and structure of the network. Specifically:

\begin{itemize}
    \item The edge weight is set by the number of messages passed between the connected users.
    \item The network is constructed as an undirected graph since interactions in the Telegram bot occur both directions.
\end{itemize}

\subsubsection{Techniques Applied in Building the Interaction Network}

Python and the NetworkX tool help to build the network programmatically. There are essentially two steps involved:

\paragraph{\textbf{Node Addition and Graph Initialization}}
The network construction begins by initializing a new undirected graph, denoted as $G$. Each user from the dataset is added to the graph as a node, with their self-reported Myers-Briggs Type Indicator assigned as a node attribute.

\paragraph{\textbf{Edge Formation and Weight Assignment}}
Edges in the graph are defined based on interactions between users. Each edge is assigned a weight corresponding to the number of messages exchanged between the two connected users.

\paragraph{Network Metrics Analysis}
To assess the structure and connectivity of the resulting network, several core metrics are computed. The \textbf{number of nodes and edges} reflects the overall size of the network. \textbf{Edge density} is calculated to determine how interconnected the network is. The \textbf{average clustering coefficient} captures the tendency of nodes to form tightly-knit communities. The number of \textbf{connected components} indicates how many isolated subgraphs exist within the network. Additionally, \textbf{degree centrality} identifies users with the highest number of direct connections, while \textbf{betweenness centrality} reveals users who act as critical bridges within the network. Finally, \textbf{closeness centrality} measures how efficiently each user can reach all other users, providing insight into information flow efficiency.

\paragraph{Homophily and Heterophily Analysis}
To examine personality-driven clustering within the network:
The analysis includes several key measures. Myers-Briggs Type Indicator homophily is computed by measuring the frequency of interactions between users with the same Myers-Briggs Type Indicator type. In contrast, Myers-Briggs Type Indicator heterophily is assessed by identifying the most common cross-type interactions. Additionally, gender homophily is analyzed by examining how often users of the same gender interact, while gender heterophily explores the prevalence of interactions between users of different genders. Finally, assortativity coefficients are calculated to quantify the extent to which nodes with similar attributes—both in terms of Myers-Briggs Type Indicator and gender—tend to be connected within the network.

\paragraph{Degree Distribution Analysis}
Whether the network exhibits scale-free characteristics by following a power-law distribution is investigated by means of the analysis of node degrees. A logarithmic scale helps one to visualize the degree distribution and evaluate its fit to theoretical models.

\paragraph{Reproducibility and Storeability}
The computed network metrics are kept in a structured CSV file for additional analysis and validation, so guaranteeing repeatability of results and facilitating comparison between several datasets.

This network representation offers a quantitative framework for comprehending how interactions motivated by personality shape social systems in anonymous digital environments.

\subsection{Calculation of Network Metrics}

To analyze the structural properties of the Myers-Briggs Type Indicator-driven interaction network, we employ a set of network metrics that provide insights into node importance, connectivity patterns, homophily, and clustering tendencies. The computations are based on graph-theoretic principles and are implemented using the \texttt{networkx} Python library.

\subsection{Centrality Measures}

Centrality measures quantify the relative importance of nodes within the network. The following centrality metrics are computed:

\subsubsection{Degree Centrality}

Degree centrality measures the number of direct connections a node has. It is computed as:

\begin{equation}
    C_D(v) = \frac{\deg(v)}{N-1}
\end{equation}

where \( \deg(v) \) denotes the degree of node \( v \), and \( N \) is the total number of nodes. The degree distribution follows a power-law pattern, indicating the presence of hubs.

\subsubsection{Betweenness Centrality}

Betweenness centrality quantifies the extent to which a node lies on the shortest paths between other nodes, acting as a bridge in the network. It is defined as:

\begin{equation}
    C_B(v) = \sum_{s \neq v \neq t} \frac{\sigma_{st}(v)}{\sigma_{st}}
\end{equation}

where \( \sigma_{st} \) is the total number of shortest paths between nodes \( s \) and \( t \), and \( \sigma_{st}(v) \) is the number of these paths that pass through \( v \). High betweenness centrality suggests that a node facilitates communication across otherwise disconnected regions.

\subsubsection{Closeness Centrality}

Closeness centrality measures how efficiently a node can reach all other nodes in the network:

\begin{equation}
    C_C(v) = \frac{N-1}{\sum_{u \neq v} d(v, u)}
\end{equation}

where \( d(v, u) \) is the shortest path between nodes \( v \) and \( u \). Nodes with higher closeness centrality are more structurally central and can disseminate information more efficiently.

\subsubsection{Eigenvector Centrality}

Eigenvector centrality extends degree centrality by considering the importance of a node’s neighbors:

\begin{equation}
    x_v = \frac{1}{\lambda} \sum_{u \in N(v)} x_u
\end{equation}

where \( N(v) \) represents the set of neighbors of node \( v \), and \( \lambda \) is the largest eigenvalue of the adjacency matrix. High eigenvector centrality indicates a node's influence within highly connected regions.

\subsubsection{Katz Centrality}

Katz centrality generalizes eigenvector centrality by incorporating both direct and indirect connections, allowing the influence of a node to propagate through the network:

\begin{equation}
    C_K(v) = \alpha \sum_{u \in N(v)} C_K(u) + \beta
\end{equation}

where \( \alpha \) is a decay factor (typically \( \alpha < \frac{1}{\lambda_{\max}} \) to ensure convergence), and \( \beta \) is an external bias term ensuring that nodes with no incoming connections still have some centrality. This measure is particularly useful for detecting the relative importance of less-connected nodes.
\subsubsection{Rich Club Coefficient}

The rich club coefficient measures how likely it is for high-degree nodes to form subgroups that are tightly connected to each other. This shows how the network is structured hierarchically. It is defined as:

\begin{equation}
    \phi(k) = \frac{2E_{>k}}{N_{>k} (N_{>k} - 1)}
\end{equation}
\noindent
where \( E_{>k} \) represents the number of edges among nodes with a degree greater than \( k \), and \( N_{>k} \) is the number of nodes with degree greater than \( k \). A higher \(\phi(k)\) value suggests that high-degree nodes preferentially connect with each other, forming a dominant core in the network. This metric is particularly useful for identifying influential clusters and understanding the hierarchical organization of social structures in personality-driven interactions.

\subsection{Assortativity and Homophily Measurements}

Assortativity quantifies the tendency of nodes to connect with similar nodes based on attributes such as Myers-Briggs Type Indicator personality type and gender. 

\subsubsection{Assortativity Coefficient}

The assortativity coefficient \( r \) measures the correlation of attributes between connected nodes:

\begin{equation}
    r = \frac{\sum_{xy} xy(e_{xy} - a_x a_y)}{\sigma_a^2}
\end{equation}

where \( e_{xy} \) is the fraction of edges connecting nodes of types \( x \) and \( y \), \( a_x \) is the fraction of nodes of type \( x \), and \( \sigma_a^2 \) is the variance of the attribute distribution. Values of \( r \) close to 1 indicate strong homophily, while values near -1 suggest heterophily.

\subsubsection{Homophily Index}

The following formula computes the proportion of interactions between users of the same Myers-Briggs Type Indicator type:

\begin{equation}
    H_{\text{Myers-Briggs Type Indicator}} = \frac{\sum_{i} E_{ii}}{\sum_{i} E_{i}}
\end{equation}

where \( E_{ii} \) represents the number of interactions between users of the same Myers-Briggs Type Indicator type, and \( E_{i} \) is the total number of interactions involving Myers-Briggs Type Indicator type \( i \).

Similarly, gender homophily is assessed as:

\begin{equation}
    H_{\text{gender}} = \frac{E_{\text{same-gender}}}{E_{\text{total}}}
\end{equation}

where \( E_{\text{same-gender}} \) represents interactions between users of the same gender.

\subsection{Clustering Coefficients and Modularity Analysis}

The clustering coefficient captures the local cohesiveness of the network, indicating the probability of two connected nodes sharing a common neighbor.

\subsubsection{Clustering Coefficient}

The local clustering coefficient for a node \( v \) is computed as:

\begin{equation}
    C(v) = \frac{2T_v}{\deg(v)(\deg(v) - 1)}
\end{equation}

where \( T_v \) is the number of triangles that include node \( v \). The global clustering coefficient is computed as:

\begin{equation}
    C = \frac{1}{N} \sum_{v} C(v)
\end{equation}

\subsubsection{Modularity}

Modularity \( Q \) measures the strength of community structure:

\begin{equation}
    Q = \frac{1}{2m} \sum_{ij} \left[ A_{ij} - \frac{k_i k_j}{2m} \right] \delta(c_i, c_j)
\end{equation}

where \( A_{ij} \) is the adjacency matrix, \( k_i \) and \( k_j \) are the degrees of nodes \( i \) and \( j \), and \( \delta(c_i, c_j) \) is 1 if nodes \( i \) and \( j \) belong to the same community. Higher modularity values indicate well-defined communities.

\subsection{Community Structure (Louvain Method)}

To identify cohesive communities within the Myers-Briggs Type Indicator-driven network, we employ the Louvain method, an optimization-based approach for detecting hierarchical modular structures. The algorithm maximizes modularity by iteratively merging nodes into communities that yield the highest modularity gain. The Louvain modularity function is computed as:

\begin{equation}
    Q = \frac{1}{2m} \sum_{i,j} \left[ A_{ij} - \frac{k_i k_j}{2m} \right] \delta(c_i, c_j)
\end{equation}

where:
- \( A_{ij} \) represents the adjacency matrix,
- \( k_i \) and \( k_j \) denote the degrees of nodes \( i \) and \( j \),
- \( \delta(c_i, c_j) \) is 1 if nodes \( i \) and \( j \) belong to the same community.

The Louvain technique repeatedly refines community assignments until no more modularity improvement is feasible. The resulting partitions expose Myers-Briggs Type Indicator personalities' underlying structural organization of interactions.
\noindent
 Following the development of the interaction network and computation of important measures, the results provide a strong basis for investigating private environment personality-driven social behavior. We quantify network connectivity, centrality distributions, homophily patterns, and modular structures using graph-theoretic approaches. This demonstrates how people's personalities shape their online interactions. By means of assortativity analysis, clustering coefficients, and the Louvain method to identify communities together, we can investigate closely the formation of Myers-Briggs Type Indicator-based networks. These methodological steps create a strict framework for assessing the impact of psychological characteristics on digital social structures, so enabling additional empirical research and interpretation of results.

\section{Results}

\subsection{General Network Statistics}
\begin{figure}[H]
\centering
\includegraphics[width=10 cm]{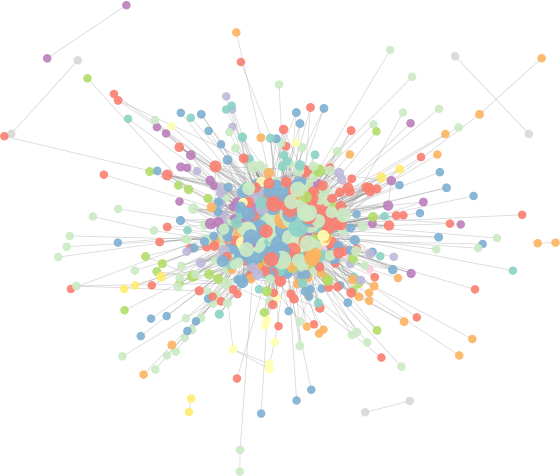}
\centering
\caption{Social Network Graph Colored by Community\label{social_network_graph}}
\end{figure}  
\noindent
The produced interaction network has \textbf{10,537} edges and \textbf{940} nodes. Calculated as \textbf{0.0239}, the edge density of the network shows a rather limited connectivity structure. The low edge density implies that the network is not densely connected since most of the nodes are connected to only a small fraction of other nodes. Several highly connected hub nodes balance the sparsity of the network, thus maintaining the network's rather good connectivity even with its sparse character.

There are \textbf{7} connected components in total, thus even if most users are part of a big connected component, some remain isolated or concentrated in small groups. Several components reflect the social fragmentation brought about by insufficient integration of specific users or subgroups into the larger network. This phenomenon is typical in real social networks, where isolated individuals or niche communities coexist with a large, linked core. Still, the largest component consists of most of the nodes, implying that the majority of users belong to one dominant interaction network.

Considering an average degree of \textbf{22.42}, every user interacts with about 22 other users on average. This statistic shows modest connectivity but also hides notable degree distribution variability. The network exhibits a highly skewed degree distribution, in which most nodes have relatively few connections, while a small number of hub nodes have significantly high connectivity. These hubs are essential for preserving network cohesiveness since they act as central points through which many interactions are directed.

The maximum degree noted at \textbf{234} suggests the presence of highly connected hub users who are fundamental in enabling interactions. These hub nodes form the structural backbone of the network, acting as conduits for communication and information flow between otherwise distant parts of the network. The prominence of hub nodes contributes to the scale-free nature of the network, where the degree distribution follows a power-law pattern. This phenomenon aligns with the principles of preferential attachment, where popular or influential nodes continue to attract new connections over time, reinforcing their centrality.

Moreover, the presence of such high-degree nodes underscores the potential vulnerability of the network to targeted attacks or failures. The removal of these hub nodes could significantly fragment the network, causing a dramatic reduction in connectivity and potentially isolating numerous subcomponents. Thus, the network's resilience heavily relies on the structural integrity of its most connected nodes.

In terms of social dynamics, hub nodes often represent influential or central users within the anonymous social network. These users may possess characteristics or behavioral patterns that make them attractive interaction partners, leading to their elevated status within the network topology. Examining the specific attributes and roles of these high-degree nodes could yield deeper insights into the social mechanisms driving connectivity and community formation within the network.

The observed degree distribution and the critical role of hub nodes emphasize the core-periphery structure characteristic of many social networks. In such structures, a densely interconnected core of influential users is surrounded by a periphery of less connected, more isolated individuals. This organizational pattern fosters efficient communication within the core while maintaining a looser connection with peripheral nodes, balancing centralized control with local autonomy.

Overall, the interaction network’s structural characteristics—marked by sparse connectivity, multiple components, and prominent hub nodes—highlight the complex interplay between local interactions and global cohesion. This configuration not only shapes information dissemination and network stability but also influences social behavior and group dynamics within the anonymous platform. Further analyses focusing on the roles and attributes of hub nodes could provide valuable insights into the emergence of influence and connectivity patterns in anonymous social settings.
\begin{figure}[H]
\centering
\begin{adjustwidth}{-\extralength}{0cm}
\subfloat[\centering Eigenvector Centrality Distribution]{\includegraphics[width=8cm]{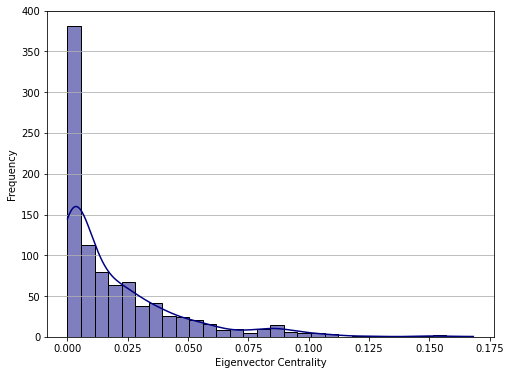}}
\hfill
\subfloat[\centering Betweenness Centrality Distribution]{\includegraphics[width=8cm]{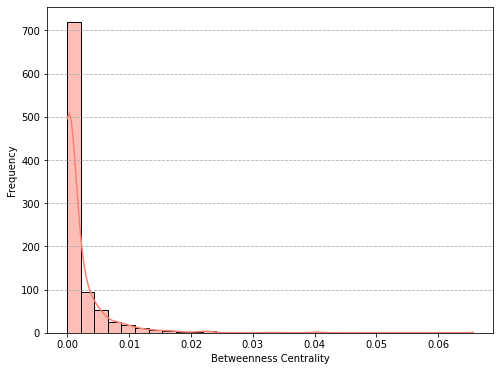}}\\
\subfloat[\centering Degree Centrality Distribution]{\includegraphics[width=8cm]{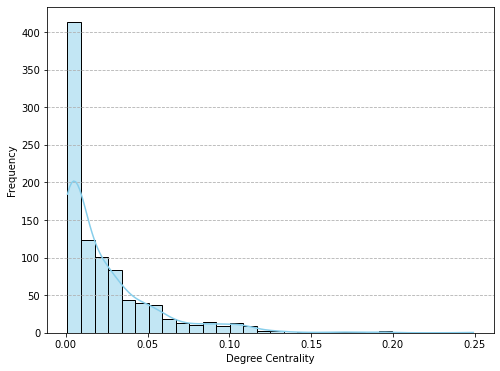}}
\hfill
\subfloat[\centering Closeness Centrality Distribution]{\includegraphics[width=8cm]{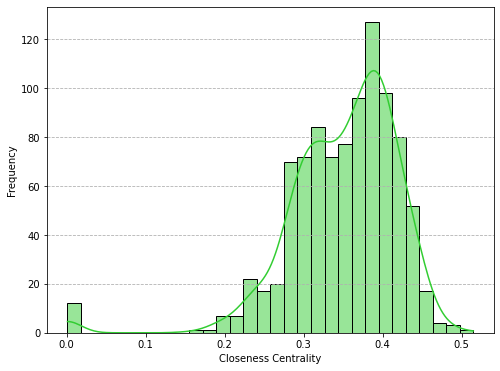}}
\end{adjustwidth}
\caption{Network's centrality distributions. (\textbf{a}) According to the eigenvector centrality distribution, most nodes remain weakly connected while a small subset of them show great influence. (\textbf{b}) The long-tail pattern of the betweenness centrality distribution suggests that only few nodes serve as important bridges between communities. (\textbf{c}) The degree centrality distribution shows a similar slanted trend, so supporting the presence of hub nodes. (\textbf{d}) The more usually distributed closeness centrality distribution indicates a more balanced reachability over the network. These distributions reveal structural relevance of nodes in the interaction network.}
\label{fig:centrality-distributions}
\end{figure}

\subsection{Centrality Measures}

\textbf{Degree Centrality:} The degree centrality distribution exhibits a skewed pattern, where a minority of nodes display exceptionally high degrees while the majority maintain a limited number of connections. This phenomenon is characteristic of scale-free networks, in which hub nodes dominate network connectivity. The presence of a node with \textbf{234} connections supports the notion that a few users play a crucial role in maintaining overall connectivity.

\textbf{Betweenness Centrality:} The mean betweenness centrality of the network is \textbf{0.0019}, indicating that most users do not act as essential bridges for information flow. However, the existence of high-degree nodes highlights the presence of a few influential users who serve as connectors between distant parts of the network, thus significantly impacting interaction dynamics.

\textbf{Closeness Centrality:} With an average closeness centrality of \textbf{0.3505}, the network demonstrates that users within sub-networks are relatively well-connected. This measure of information transfer efficiency indicates that most users can reach others through relatively few intermediate steps.

\textbf{Eigenvector Centrality:} The mean eigenvector centrality of \textbf{0.0201} indicates that influential nodes often cluster together, forming structurally significant subgroups. This pattern underscores how central figures shape the interaction dynamics within the network.

\begin{figure}[H]
\centering
\begin{adjustwidth}{-\extralength}{0cm}
\subfloat[\centering Rich-Club Coefficient by Degree]{\includegraphics[width=8cm]{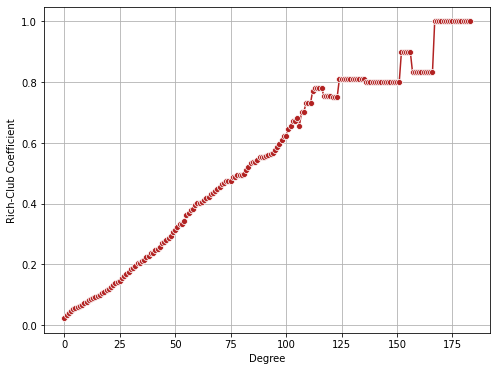}}
\hfill
\subfloat[\centering Clustering Coefficient by Degree]{\includegraphics[width=8cm]{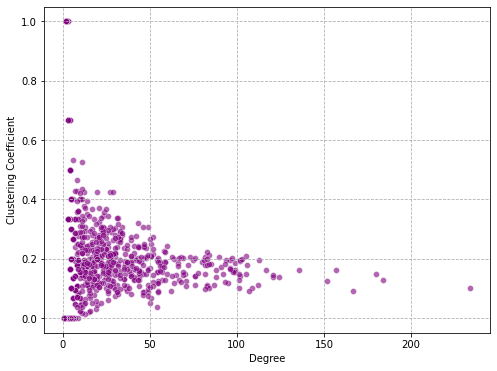}}
\end{adjustwidth}
\caption{Analysis of degree-based network characteristics. (\textbf{a}) The positive correlation shown by the rich-club coefficient by degree indicates that nodes with higher degrees typically create tightly connected subgroups. The coefficient approaches 1 as the degree rises, implying that high-degree nodes preferably interact with each other to generate a network with a top-notional core structure. (\textbf{b}) The clustering coefficient by degree follows an inverse trend: low-degree nodes show greater clustering while high-degree nodes typically show lower clustering coefficient. This is unique of scale-free networks, in which hubs link several communities instead of creating close clusters.
}
\label{fig:degree-based-metrics}
\end{figure}

\subsection{Network Structure and Clustering}

\textbf{Clustering Coefficient:} With an average clustering coefficient of \textbf{0.1566}, some users participate in tightly linked clusters. This suggests that although keeping a general sparse structure, the network shows localised pockets of high connectivity.

\textbf{Rich Club Coefficient:} High-degree users preferably connect with each other at a somewhat higher rate than expected under random conditions, according to the rich club coefficient at a maximum degree greater than 5, \textbf{1.0058}. This supports the existence of network core-periphery structures.

\begin{figure}[H]
\centering
\includegraphics[width=10cm]{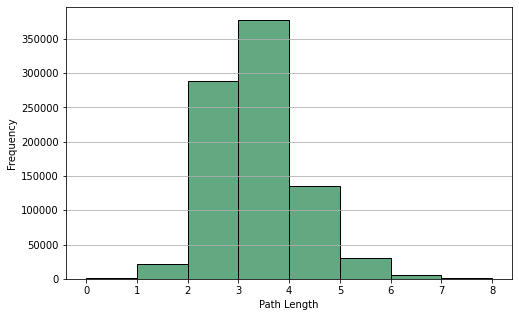}
\caption{Shortest Path Length Distribution. The histogram illustrates the distribution of shortest path lengths between all pairs of connected users in the network. The majority of paths fall between lengths 3 and 4, indicating that most users can reach each other through relatively short chains of interactions. This structure reflects high overall network navigability and supports the presence of small-world characteristics, where most nodes are just a few steps apart despite the network’s size and sparsity.}
\label{fig:shortest-path}
\end{figure}

\subsection{Shortest Path Length Distribution}

In network analysis, the shortest path length distribution is an important statistic since it measures the minimum number of edges needed to go between any two connected nodes. Calculated shortest path lengths for every pair of users in the network are displayed in Figure~\ref{fig:shortest-path}. The great majority of users can reach each other in just three or four interaction steps, as the distribution shows a sharp peak between path lengths of three and four. This realization emphasizes the amazing navigability and effective communication channels of the network.

Short path lengths are rather common in line with the well-known \textit{small-world phenomenon}, in which most nodes are only a few hops apart despite the rather large network size and structural sparsity. This feature of social networks suggests that knowledge and influence might spread quickly across the system. Short paths improve communication efficiency and help to build strong connectivity so that far-off nodes may remain reachable with few intermediate steps.

Moreover, longer path lengths—albeit in rather lower frequencies—indicate the presence of peripheral or marginal nodes only indirectly connected to the core of the network. These nodes might be users who belong to isolated subcommunities with less connection to the main core structure or who interact seldom.

Such a distribution of path lengths supports the idea of a \textit{core-periphery} structure, in which strongly connected hubs act as central nodes allowing communication between several subgroups. The network's short average path length indicates that users may easily access knowledge from other areas of the network with minimum effort, so indicating its efficient navigability. Even in an anonymous interaction environment, this structural feature helps to preserve a high degree of social cohesiveness and enable the fast spread of knowledge.

This study is essential to the characterization of interaction patterns inside the anonymous social network since the understanding of the efficiency and resilience of network topology depends on the shortest path lengths. Consequently, it is most fitting to add this subsection right after the analysis of clustering coefficients and rich-club characteristics in the \textbf{Network Structure and Clustering} part of the paper. This arrangement guarantees a rational development from local connectivity measures to global navigability characteristics.

\begin{figure}[H]
\centering
\begin{adjustwidth}{-\extralength}{0cm}
\subfloat[\centering Myers-Briggs Type Indicator Homophily Rate by Type]{\includegraphics[width=8cm]{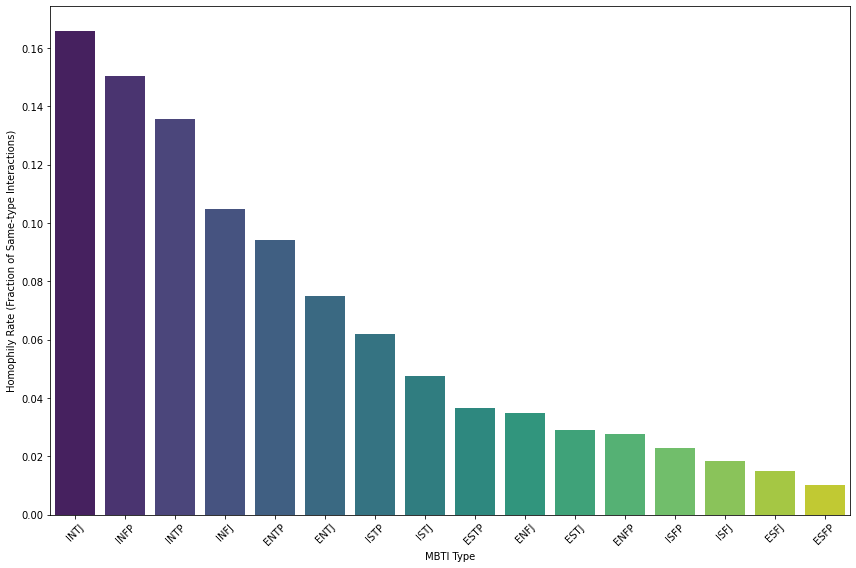}}
\hfill
\subfloat[\centering Top Myers-Briggs Type Indicator Heterophilous Interactions]{\includegraphics[width=8cm]{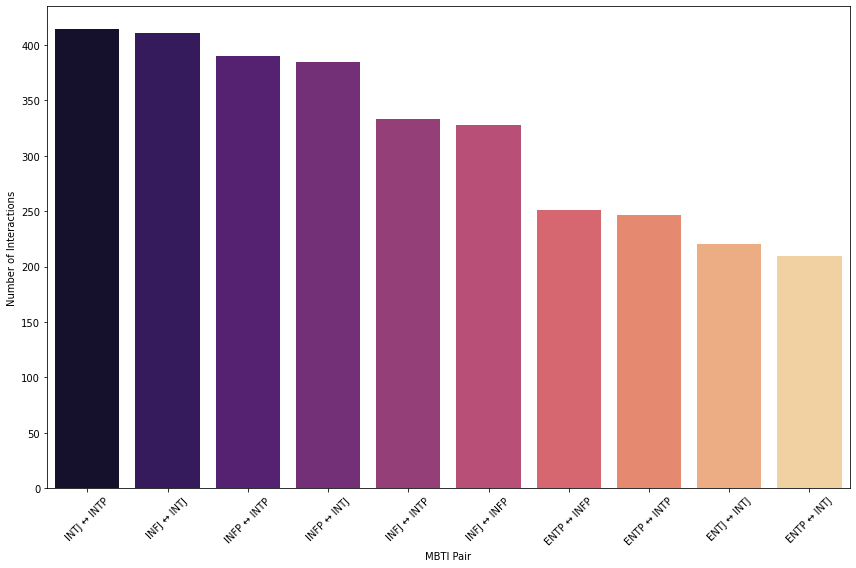}}
\end{adjustwidth}
\caption{Myers-Briggs Type Indicator-based interaction patterns. (\textbf{a}) The Myers-Briggs Type Indicator homophily rate by type shows the proportion of interactions occurring between users with the same Myers-Briggs Type Indicator type. The INFJ, INFP, and INTJ types exhibit the highest levels of homophily, indicating a strong preference for interacting within their personality groups. (\textbf{b}) The most frequent Myers-Briggs Type Indicator heterophilous interactions reveal cross-type communication patterns. The most common pairings include intuitive-thinking (NT) and intuitive-feeling (NF) types, suggesting that these personality traits may foster engaging and dynamic interactions in anonymous settings.}
\label{fig:Myers-Briggs Type Indicator-interactions}
\end{figure}
\subsection{Myers-Briggs Type Indicator Homophily and Heterophily Analysis}

\textbf{Myers-Briggs Type Indicator Homophily:} Myers-Briggs Type Indicator personality types have a homophily rate calculated as \textbf{0.1068}, hence about \textbf{10.7\% of all interactions} take place between users with the same personality type. This implies that, even if people might show some inclination for interacting with like-minded personalities, the network stays mostly heterophilous in terms of Myers-Briggs Type Indicator-driven interactions.

Among the personality types, homophilic tendencies show rather different variation. With their highest homophily rates—\textbf{INTJ (16.6\%)}, \textbf{INFP (15.0\%)}, \textbf{INTP (13.6\%)}, and \textbf{INFJ (10.5\%)}, these types are most likely to participate in self-reinforcing social circles. The lowest homophily rates among \textbf{ESFP (1.0\%)}, \textbf{ESFJ (1.5\%)}, and \textbf{ISFJ (1.8\%)}, which imply that these personality types interact more often with different groups than their own.

Myers-Briggs Type Indicator heterophily, on the other hand, explains \textbf{89.3\%} of all interactions and shows a strong inclination toward cross-type involvement. The most often occurring heterophilous interactions in an anonymous environment are those between intuitive-thinking (NT) and intuitive-feeling (NF) personality types. This implies that persons with complementary cognitive styles are more likely to be involved. The referenced heatmap figure shows a graphic depiction of these interaction dynamics, emphasizing Myers-Briggs Type Indicator interaction patterns both strongest and weakest.

\subsection{Gender Homophily and Heterophily Analysis}

\textbf{Gender Homophily:} Gender-based interaction analysis exposes clear variations in homophilic behavior. Computed homophily rates for every gender are:
\begin{itemize}
    \item \textbf{Female homophily rate}: 55.6\%
    \item \textbf{Male homophily rate}: 23.4\%
\end{itemize}
These findings show that whereas male users show a considerably higher inclination toward cross-gender interaction, female users show a stronger inclination for interacting inside their gender group. 

\begin{figure}[H]
\centering
\includegraphics[width=8cm]{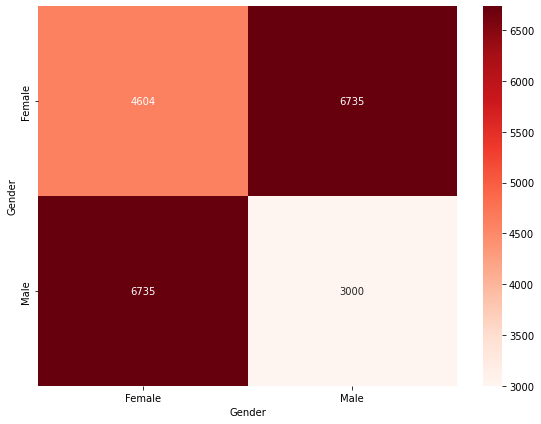}
\caption{Symmetric Gender Interaction Heatmap. This heat map shows male and female users' interaction distribution. The great intra-female interaction count shows that female users show a stronger inclination for interacting inside their gender group. Conversely, male users interact with people of other sexes far more frequently, which supports the hypothesis that male-driven interactions in the dataset are typically not friendly.}
\label{fig:gender-heatmap}
\end{figure}

\textbf{Gender Heterophily:} The dataset mostly consists of cross-gender interactions; male-female interactions are rather more frequent. Male users are more likely to engage in heterophilous activity, as the total number of male-female and female-male interactions much exceeds same-gender interactions. Emphasizing the distribution of interactions both between and inside gender groups, the \ref{fig:gender-heatmap} shows this trend.

\textbf{Interpretation of Findings:} These results fit accepted behavioral theories on gender-based communication preferences, whereby male users show more general exploratory engagement while female users may gravitate toward intra-gender socializing due to perceived social affinity. Although Myers-Briggs Type Indicator-based interactions remain mostly heterophilous, gender plays a more structured role in determining social interactions; female users exhibit higher in-group preference and male users engage more randomly across genders.

\begin{figure}[H]
\centering
\includegraphics[width=12cm]{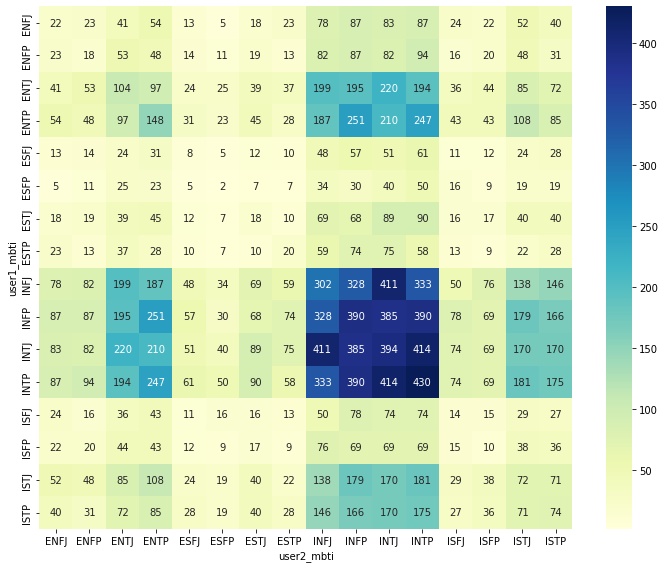}
\caption{Myers-Briggs Type Indicator Interaction Heatmap. The interaction frequencies between many Myers-Briggs Type Indicator personality types are shown on the heatmap. Darker tones indicate more interaction counts; intuitive-thinking (NT) and intuitive-feeling (NF) types have most frequent interactions. Particularly between the NT and NF groups, the clustering pattern reveals a great degree of heterophilous interaction. Conversely, there are less interactions both inside and between the groups of sensing-feeling (SF).}
\label{fig:Myers-Briggs Type Indicator-heatmap}
\end{figure}

\subsection{Degree Distribution and Power Law Fit}
\begin{figure}[H]
\centering
\includegraphics[width=8.86cm]{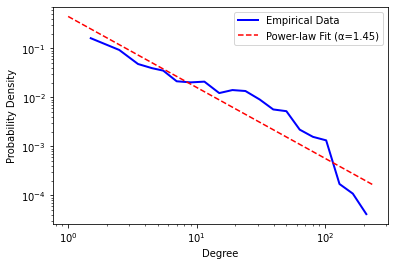}
\caption{Degree Distribution and Power-Law Fit. The plot visualizes the degree distribution of the network on a log-log scale, comparing the empirical data (blue) with a fitted power-law model (red dashed line). The fitted exponent (\(\alpha = 1.45\)) suggests that the network follows a scale-free structure, where a few high-degree nodes (hubs) dominate connectivity, while most nodes maintain relatively low-degree connections. This pattern is characteristic of many real-world social networks, reinforcing the presence of influential users within the interaction structure.}
\label{fig:power-law}
\end{figure}

Figure~\ref{fig:power-law} shows that the network's degree distribution follows a power-law form with an exponent \( \alpha = \textbf{1.45}\). Comparing the empirical degree distribution (blue line) with a fitted power-law model (red dashed line) the log-log plot reveals that the network has a scale-free structure. The results show that although most nodes retain lesser degrees, a small number of high-degree hub nodes are central for connectivity.

Common to self-organized complex systems and extensively noted in many social and communication networks is this kind of distribution. Hubs imply that interactions arise from preferential attachment dynamics, in which highly connected nodes attract more connections over time rather than from random distribution. Information flow, resilience, and community building inside the network depend much on this structural feature.

\begin{figure}[H]
\centering
\includegraphics[width=8.86cm]{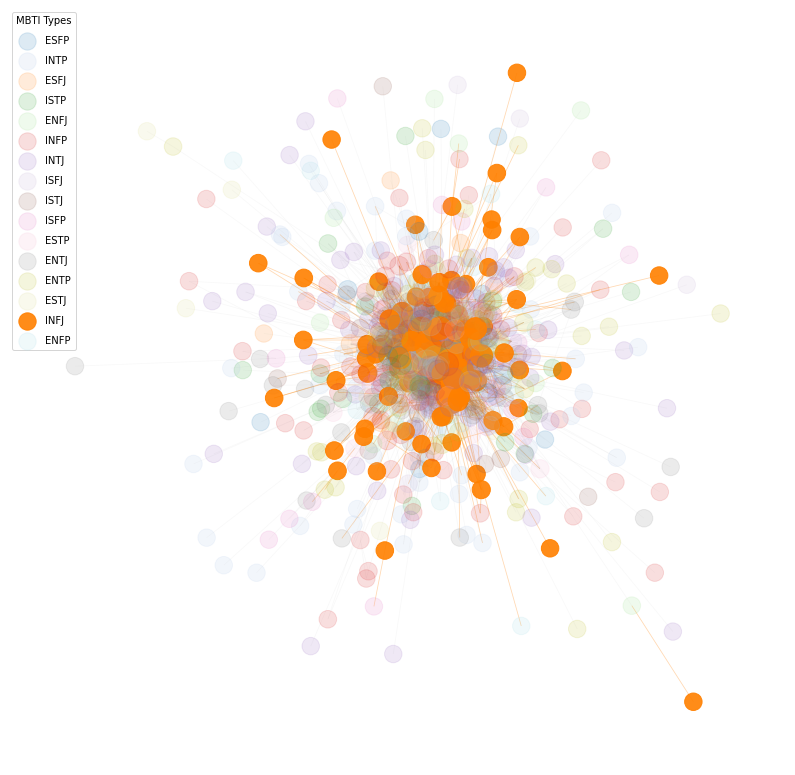}
\caption{a view of Myers-Briggs Type Indicator-colored network community with largest concentration. Emphasizing the distribution of INFJ users (orange) among other Myers-Briggs Type Indicator types (lighter colors), the figure shows the framework of the biggest found community. According to the network, INFJ users are central and well-integrated, generating homophilous—same-type—as well as heterophilous—cross-type interactions. Different Myers-Briggs Type Indicator types in the community point to the fact that personality shapes social interactions but does not produce exactly segregated groups.
}

\label{fig:infj-network}
\end{figure}

\subsection{Community Detection}

\textbf{Number of Communities:} Considering that interactions are arranged in separate groups, the Louvain method detects \textbf{32} communities inside the network as shown in \ref{social_network_graph}. The recorded modularity score of \textbf{0.2584} shows that the network is not highly modular even if community structures exist.

\textbf{Myers-Briggs Type Indicator and Community Formation:} As shown by \ref{fig:infj-network} More investigation reveals that the largest community consists mostly of users with the \textbf{INFJ} personality type, implying that some personality groups might be more likely to cluster together. Myers-Briggs Type Indicator assortativity is still low overall, thus personality does not seem to be a major determinant of community formation.

\subsection{Interaction Dynamics}

\textbf{Frequent Myers-Briggs Type Indicator Pairings:} While the least frequent interaction is seen between \textbf{ENFJ and ESFP}, the most often occurring Myers-Briggs Type Indicator-based interaction is between \textbf{INFP and INTP}. This outcome implies that, perhaps by means of shared cognitive styles or conversational tendencies, intuitive-thinking and intuitive-feeling personalities show stronger affinities in anonymous interactions.

\textbf{Heterophilic Patterns:} The great frequency of heterophilic interactions implies that users often interact across personality boundaries, so supporting the theory that anonymity lowers conventional social constraints. Fascinating is it that some groups exhibit dominant Myers-Briggs Type Indicator clusters. This implies that self-organized structures may still develop through interpersonal interactions even if people do not actively seek for people with similar personalities.

\textbf{Bridge Personalities:} Some Myers-Briggs Type Indicators—especially those with high eigenvector and betweenness centrality scores—function as links between several communities. Such bridge personalities help to promote cross-group communication and network cohesiveness, so supporting structural hub function in preserving network integrity.
\begin{figure}[H]
\centering
\includegraphics[width=8.86cm]{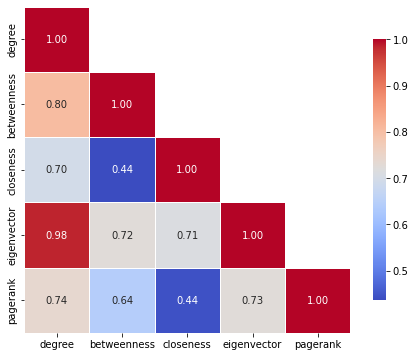}
\caption{Correlation Heatmap of Centrality Measures. The heatmap shows the pairwise Pearson correlation coefficients among five key centrality metrics: degree, betweenness, closeness, eigenvector, and PageRank. Strong positive correlations are observed between \textbf{degree and eigenvector centrality} (\(r = 0.98\)) and between \textbf{degree and betweenness} (\(r = 0.80\)), indicating that nodes with more connections tend to also be more influential within the network’s structure. In contrast, \textbf{closeness centrality} shows weaker correlations with the other metrics, suggesting it captures a different dimension of network centrality focused on distance-based accessibility.}
\label{fig:centrality-correlation}
\end{figure}
We computed pairwise Pearson's correlation coefficients among five centrality measures: degree, betweenness, closeness, eigenvector, and PageRank to better grasp the interactions between several notions of importance in the network. The degree and eigenvector centrality (\(r = 0.98\) have a strong positive correlation as shown in Figure~\ref{fig:centrality-correlation}, suggesting that highly connected nodes are typically also influential in terms of their connections to other central nodes. Likewise, degree is favorably correlated with betweenness (\(r = 0.80\) and PageRank (\(r = 0.74\), implying that nodes with more connections are more likely to be bridges or rank highly in terms of influence propagation.

In contrast, closeness centrality shows notably weaker correlations with the other metrics—most notably with PageRank (\(r = 0.44\)) and betweenness (\(r = 0.44\))—indicating that it captures a more distinct aspect of node centrality based on the average distance to all other nodes in the network. These results confirm that while several centrality measures overlap conceptually, they also capture unique dimensions of structural prominence, reinforcing the value of a multi-metric approach to identifying key actors in the network.

\section{Summary of Findings}

Myers-Briggs Type Indicator-driven interactions in a large-scale anonymous chat network are analysed to provide several interesting new perspectives on the dynamics and structure of personality-based social behaviour in identity-free digital environments.
Comprising 940 nodes and 10,537 edges, the built network shows classic features of a \textit{scale-free system}: a highly skewed degree distribution with a power-law exponent \( \alpha = 1.45 \). This implies the existence of preferential attachment mechanisms, in which a small subset of highly connected users (hubs) disproportionately influence the information flow and social connectivity. 

While the existence of a weak but present \textit{rich-club effect} (\( \phi(k) > 1 \) highlights a hierarchical structure in which central personalities form a loosely knit elite core, the average clustering coefficient (\( C = 0.1566 \) indicates moderate local cohesion. These trends correspond with theoretical expectations in self-organizing social systems, in which structural redundancy coexists with vulnerability to targeted elimination and influence is asymmetrically distributed.

Though faintly, traces of homophily driven by personality remain even on the anonymous platform. The general Myers-Briggs Type Indicator homophily rate (\( H_{\text{Myers-Briggs Type Indicator}} = 0.1068 \)) validates a modest inclination of users to interact with others of the same personality type. Particularly introverted intuitive types including \textbf{INTJ}, \textbf{INFP}, and \textbf{INFJ} show stronger self-affinity tendencies. These kinds might be drawn toward conversational rhythms they know or toward ideologically coherent debates.

On the other hand, extraverted sensing-feeling types (\textbf{ESFP}, \textbf{ESFJ}) show minimal homophily, reflecting greater openness to diverse interactions—a result compatible with their behavioral typologies stressing adaptation and social harmony.

With notable frequency among pairings like \textbf{INFP--INTP} and other NT--NF dyads, roughly \textbf{89.3\%} of all interactions are heterophilous in character. These findings strongly support the \textit{homophily amplification hypothesis}: they imply that the anonymity of the platform promotes exploratory interaction across personality boundaries.

Fascinatingly, heterophilous interaction is not random but rather follows clear trends. Cross-type interactions are dominated by intuitive types (N), presumably because of shared abstract thinking patterns that go across Myers-Briggs Type Indicator variations. These results suggest that cognitive compatibility might overcome surface-level personality difference to enable interesting communication.

In interaction dynamics, gender plays a rather more ordered role. Female users show a higher gender homophily rate (55.6\%) than men (23.4\%), implying intra-gender bonding among women, maybe resulting from common emotional or conversational rules. Conversely, male users show a clear heterophilic inclination and interact more often between gender lines.

This asymmetry conforms with earlier studies on conversational preferences in anonymous digital environments and reflects sociopsychological results on gender-based communication styles. It emphasizes even in identity-concealed settings the fact that \textbf{gender-based clustering remains robust} while personality homophily is subtle.

Using the Louvain algorithm, community detection reveals 32 discrete clusters with a modest modularity score (\( Q = 0.2584 \%)\), so indicating the existence of latent substructures without exact community separation. Myers-Briggs Type Indicator clusters—especially those centered on INFJ users—suggestive of subtle ways in which personality may still influence community coalescence.

Often from intermediate Myers-Briggs Type Indicator types (e.g., INTP, ENFP), users with high eigenvector and betweenness centrality behave as \textit{bridge personalities}, so promoting cross-community integration and network stabilization. These people occupy structurally important roles that link otherwise far-off personality clusters and advances information diffusion.

All taken together, the results show a complex interaction between structural features of digital networks and psychological aspects. Myers-Briggs Type Indicator-based homophily exists, but it is eclipsed by heterophilous exploration most likely driven by the affordances of anonymity and the asynchronous character of text communication.

Rich-club effects, self-organized clusters, and the function of bridge personalities point to emergent personality-driven structures free from explicit identity disclosure. This highlights how subtly cognitive and communicative style shapes social topology—even in cases when conventional identity signals are eliminated.

\subsection{Theoretical Implications}

This study contributes to both psychology and network science by:

\begin{itemize}
    \item Demonstrating that personality traits can subtly shape interaction structures in anonymous digital environments.
    \item Providing empirical evidence that heterophily may dominate in high-anonymity, low-friction networks, particularly among cognitively compatible personality types.
    \item Highlighting that gender remains a robust axis of social structuring, even in anonymous spaces, potentially due to deeper socio-cognitive scripts.
    \item Validating the presence of scale-free, modular, and bridge-mediated network architectures arising from voluntary social engagement patterns.
\end{itemize}

These discoveries guide more general theoretical models of digital sociality by implying that the psychological blueprint of people shows through emergent network dynamics even in the absence of identity markers.

\subsection{Practical Relevance}

The outcomes affect digital community-building techniques, recommendation systems, and anonymous platform design. Knowing how personality shapes anonymous interaction patterns helps one guide treatments meant to enhance user experience, encourage positive participation, and create algorithms respecting individual differences without depending on identity disclosure.

\section{Discussion and Future Work}

\subsection{Integrating Personality Psychology with Network Science}

This study offers empirical evidence showing clear traces of psychological traits—especially Myers-Briggs Type Indicator personality types—in anonymous digital interactions. Though the general network architecture shows a high degree of heterophily, subtle homophilic patterns remain especially among introverted intuitive types such \textbf{INFJ}, \textbf{INTJ}, and \textbf{INFP}. These results imply that users may implicitly search for cognitive resonance in others even in absence of identity cues, so supporting the idea that attraction motivated by personality functions both consciously and subconsciously.

Especially between intuitive-thinking (NT) and intuitive-feeling (NF) types, the frequency of cross-type interactions supports the heterophily amplification hypothesis. These interactions might profit from the complementary character of abstract thinking and emotional expressiveness, which would allow more intense dialogues. This corresponds with earlier studies in interpersonal psychology implying that complementary qualities can improve mutual understanding and involvement in both offline and online environments.

From a network-theoretic point of view, the scale-free topology, rich-club structure presence, and development of bridge personalities point to a self-organizing interaction ecology. These structural features imply that personality shapes not only dyadic interactions but also meso-level formations including information bridges and communities. The results showing users with particular Myers-Briggs Type Indicator profiles (e.g., INTP, ENFP) typically occupy high centrality roles emphasizes the possibility for particular personality types to act as social facilitators in anonymous environments.

\subsection{The Role of Anonymity in Modulating Social Behavior}

Anonymity brings a special vitality into social contact. One could argue that it lessens social inhibition, so enabling users to interact more freely across gender and personality lines. Conversely, it eliminates the usual social signals that support quick personality detection and alignment. According to the data, in these kinds of settings users do not specifically search for similarity; rather, emergent conversational compatibility shapes interaction more than pre-existing stereotypes or identity-based prejudices.

Weak Myers-Briggs Type Indicator assortativity and the rather low modularity score confirm that personality type by itself does not rigorously determine community structure. Instead, communities seem to develop naturally around dynamic patterns of interaction, in which case bridge personalities are essential for preserving cohesiveness. This questions conventional ideas of personality homophily by showing how more diverse and inclusive social environments might be created from identity-blind platforms.

\subsection{Limitations and Methodological Considerations}

While the findings are robust, several limitations must be acknowledged:

\begin{itemize}
    \item \textbf{Voluntary Myers-Briggs Type Indicator Self-Report:} The Myers-Briggs Type Indicator data relies on self-disclosure, introducing potential bias or inaccuracy. Although the anonymity may reduce social desirability effects, the possibility of misclassification cannot be entirely ruled out.
    
    \item \textbf{Static Snapshot Analysis:} The network is constructed based on aggregated interaction data over six months. Temporal fluctuations in user behavior, evolving community dynamics, and personality-related interaction sequences are not fully captured.
    
    \item \textbf{Context-Free Communication:} The analysis abstracts away from message content. While structural patterns provide rich insight, incorporating semantic and sentiment analysis could better contextualize the role of personality in conversational tone and topic affinity.
    
    \item \textbf{Limited Cultural and Platform Diversity:} The findings are drawn from a single platform (Telegram) with a specific demographic and usage pattern. Results may not generalize across platforms with different interaction norms, user bases, or affordances.
\end{itemize}

\subsection{Future Directions}

Building on this study, several promising avenues for future work emerge:

\begin{enumerate}
    \item \textbf{Longitudinal Network Evolution:} Monitoring network changes across time helps one to better grasp homophily and heterophily trajectories by revealing how personality-based communities arise, dissolve, or evolve.
    
    \item \textbf{Temporal Interaction Modeling:} Time-series models or sequence-based graph approaches (e.g., dynamic graphs, temporal motifs) could expose periodicities and personality-dependent interaction rhythms.
    
    \item \textbf{Semantic Enrichment:} Combining natural language processing (NLP) methods to examine message content—such as sentiment dynamics, topic modeling, emotional profiling—would help one to better understand how personality shapes communication style and emotional alignment.
    
    \item \textbf{Cross-Platform and Cross-Cultural Studies:} Comparative studies spanning cultural groups or anonymous platforms (such as Reddit or Omegle) can test the stability and universality of the noted trends.
    
    \item \textbf{Agent-Based Modeling and Simulation:} Complementing empirical results, simulating personality-driven agents in synthetic networks can assist to isolate causal mechanisms and test hypotheses under controlled conditions.
    
    \item \textbf{Mixed-Modal Personality Assessment:} Incorporating unobtrusive personality inference from behavioral data (e.g., message timing, linguistic style, emoji use) could provide a more reliable and scalable alternative to self-reported Myers-Briggs Type Indicator.
\end{enumerate}

\section{Conclusion}

This study shows in general that personality—especially as judged by Myers-Briggs Type Indicator—subtly influences social interactions and network building in anonymous online environments even in the absence of identity cues. Although heterophily is the norm, homophily pockets and personality trait-based clustering imply that digital environments appearing chaotic may really have some order to them. These results provide fresh tools and viewpoints for comprehending human behavior in the digital age, so augmenting a growing body of research bridging computational social science, personality psychology, and complex networks.

\vspace{6pt}

\informedconsent{Informed consent was obtained from all subjects involved in the study.}

\dataavailability{The dataset used in this study was collected from an anonymous Telegram social network, in full compliance with the platform’s Terms of Service. All user data were \textit{anonymized} prior to analysis to ensure the privacy and confidentiality of participants. No personally identifiable information (PII) was collected, stored, or processed at any stage. The study adheres to ethical guidelines regarding user data and digital research ethics.

Due to the sensitive nature of the conversations and to protect user anonymity, the dataset cannot be made publicly available. However, anonymized, aggregated metrics and summary statistics used for analysis may be made available upon reasonable request to the corresponding author, subject to ethical review and compliance with data protection regulations.}

\conflictsofinterest{The authors declare no conflicts of interest.} 



\reftitle{References}

\bibliography{bibliography.bib}

\isPreprints{}{}
\end{document}